\documentclass[11pt]{article}
\usepackage{textcomp}
\usepackage{graphicx}
\usepackage{wasysym}
\usepackage{float,array}

\newcommand{\Keywords}[1]{\par\noindent 
{\small{\bf Keywords\/}: #1}}

\begin{document}

\title{Reproductive isolation between phylogeographic lineages scales with divergence}
\author{Sonal Singhal and Craig Moritz \\ 
singhal@berkeley.edu \\
craig.moritz@anu.edu.au \\ \\
Museum of Vertebrate Zoology \\ University of California, Berkeley \\ 3101 Valley Life Sciences Building \\ Berkeley, California 94720-3160 \\ \\
Department of Integrative Biology \\ University of California, Berkeley \\ 1005 Valley Life Sciences Building \\ Berkeley, California 94720-3140 \\ \\
Research School of Biology \\ The Australian National University \\ Building 116 \\ Acton, ACT 0200
}
\date{}
\maketitle

\begin{abstract}
Phylogeographic studies frequently reveal multiple morphologically-cryptic lineages within species. What is yet unclear is whether such lineages represent nascent species or evolutionary ephemera. To address this question, we compare five contact zones, each of which occurs between eco-morphologically cryptic lineages of rainforest skinks from the rainforests of the Australian Wet Tropics. Although the contacts likely formed concurrently in response to Holocene expansion from glacial refugia, we estimate that the divergence times ($\tau$) of the lineage-pairs range from 3.1 to 11.5 Myr. Multilocus analyses of the contact zones yielded estimates of reproductive isolation that are tightly correlated with divergence time and, for longer-diverged lineages ($\tau >$ 5 Myr), substantial. These results show that phylogeographic splits of increasing depth can represent stages along the speciation continuum, even in the absence of overt change in ecologically relevant morphology. 
\\ 
\end{abstract}

\section{Introduction}
There is now abundant evidence for deep phylogeographic divisions within traditionally described taxa, suggesting that morphologically cryptic species are common \cite{TREEcryptic}. Indeed, deep phylogeographic structure based on mitochondrial DNA (mtDNA), and confirmed by multilocus nuclear DNA (nDNA), is increasingly used as an initial step in species delimitation via integrative taxonomy \cite{Padial}. As we grow better able to identify evolutionarily independent genetic lineages within morphologically-defined species, what is often missing is both an understanding of the forces leading to this diversity and evaluation of whether these lineages are more than ephemera \cite{RosenblumEph}. One way to start addressing these questions is to test for reproductive isolation (RI) among such morphologically cryptic lineages. Is there substantial RI between cryptic lineages, and how does this scale with divergence time and historical gene flow?  Answers to this question will inform modern systematics and contribute to our understanding of speciation processes.  

It has long been supposed that phylogeographic lineages represent a step in the continuum from population divergence to speciation \cite{Avise}. More generally, speciation theory posits that RI, especially post-zygotic RI, increases with divergence time, with the tempo and form of the relationship depending on the genetic architecture of Dobzhansky-Muller incompatibilities (DMIs) and the interaction of selection, drift, and gene flow \cite{GavriletsBook,Goubriere,Orr}. But, given this, the heterogeneity of divergent selection is expected to blur the relationship between RI and divergence \cite{GavriletsBook}. A growing body of evidence supports a general increase in RI with divergence time; however, with few exceptions \cite{Weta}, these results derive from analyses of phenotypically distinct species pairs \cite{Darters,Sasa}.  As phylogeographic lineages within morphologically defined species are often parapatrically distributed, comparative analyses of RI indices in secondary contact zones could provide a unique window into the dynamics of eco-phenotypically cryptic speciation \cite{HarrisonBook}. Such studies have the added advantage of addressing the evolution of RI in nature, in the organisms' ecological context, rather than laboratory crosses, as is more common in the literature.

To investigate the evolution of RI in nature, we exploit a system characterized by climate-driven fluctuations in habitat extent and connectivity during the Neogene -- the rainforests of the Australian Wet Tropics (AWT; Fig. 1).  Extended periods of retraction of rainforests to mesic mountain tops has resulted in pronounced phylogeographic structure within endemic faunal species, but with variable levels of sequence divergence and divergence time among intraspecific lineages \cite{Moritz2009ProcRoySocB,Bell10}.  Where tested, these mtDNA lineages are generally corroborated by multilocus nuclear gene analysis \cite{Moritz2009ProcRoySocB} (but see \cite{saproSS}); however, eco-morphological divergence is subtle or absent \cite{Bell10,Hoskin2011}. Following Holocene rainforest expansions, over twenty contact zones involving pairs of morphologically cryptic lineages formed between the historic refugia \cite{Moritz2009ProcRoySocB}.  These contact zones provide a natural experiment with which to test the hypothesis that RI increases with divergence time among cryptic lineages. Previous studies of contact zones have revealed outcomes ranging from negligible to strong RI \cite{Moritz2009ProcRoySocB}, including one case of speciation by reinforcement \cite{Hoskin2005}.  However, the cases studied to date are taxonomically and ecologically heterogeneous. 

Here, we use comparative analysis of RI within five contact zones involving lineage-pairs from a closely-related and ecologically-similar clade of terrestrial rainforest skinks, across which there are varying levels of sequence divergence among component clades: (\emph{Carlia rubrigularis} N/S, \emph{Lampropholis coggeri} N/C, \emph{L. coggeri} C/S, \emph{Saproscincus basiliscus} N/C, and \emph{S. basiliscus} N/\emph{S. lewisi}; Fig. 1).  We combine genome-scale analyses of divergence history between allopatric populations with multilocus analysis of intensively sampled contact zones to test for increasing RI with divergence time. We assume that per-generation dispersal rates are similar across lineages, which indirect dispersal estimates from this clade support \cite{Phillips2004,GilliesHZ}. Because the focal lineages are ecologically-similar, rainforest-edge species that likely tracked the expanding rainforest front closely \cite{Williams10}, we further assume that the contact zones formed concurrently. Following from these assumptions, we predict that RI scales closely with divergence, especially given the limited ecomorphological divergence of these lineage-pairs \cite{Bell10} and, for at least one lineage-pair, apparent absence of mate choice \cite{DolmanAM}. More specifically, we predict that, as divergence time increases, cline widths should narrow, clines should exhibit less variance in cline width, disequilibrium -- both within- and between-loci -- should increase, and frequency of hybrids within the hybrid zone should decrease \cite{BartonGale}.

\section{Results}
\subsection{Eco-morphological divergence}
We used morphological data, including ecologically relevant traits such as body size, limb length and head dimensions, to test for phenotypic divergence across the major phylogeographic lineages within each traditionally defined species (mean sample size, $\bar{N}$=155). We summarized the data as two principal component axes that explained over 97\% of the variation and, using these axes, found little significant morphological variation across phylogeographic lineage-pairs (Fig. S2). Where there is significant variation (Fig. S2), the differences are sex-specific and of small magnitude -- e.g., mean body size for female \emph{L. coggeri} varies from 35.7$\pm$2.7 to  38.5$\pm$3.3 mm by lineage.

\subsection{Fitting divergence histories}
Many comparative studies use genetic distance as a proxy for time since divergence \cite{Sasa,Moyle}; however, genetic distance might be decoupled from divergence time, especially if migration rates are high or ancestral population sizes are large \cite{NielsenWakeley}. Accordingly, we inferred divergence time and other demographic parameters for each lineage-pair by fitting an isolation-with-migration model to genetic data. For \emph{S. basiliscus} N and its sister species \emph{S. lewisi}, we used previously published genetic data for eight loci \cite{saproSS} to infer model parameters with {\sffamily IMa2} \cite{IMa2}. For the remaining contacts, we collected genomic data by sequencing transcriptomes from five individuals per lineage. These individuals were sampled far from the contact zone and were thus unlikely to contain recently introgressed alleles. After assembling and annotating these data, we inferred high-quality, high-coverage and non-coding single nucleotide polymorphisms (SNPs), finding an average of 30.8K SNPs per lineage-pair (Table S3). For each lineage-pair, we summarized the distribution of allele frequencies for shared and private alleles as an unfolded two-dimensional site frequency spectrum (2D-SFS)  (\cite{Nielsen2012}; Fig. S3), which we used with {\sffamily dadi} to fit a isolation-with-migration model \cite{Gutenkunst2009}.

Fitting a divergence model of isolation-with-migration to these data gave two primary results. First, divergence times vary 3.7$\times$, from 3.1 mya to 11.5 mya. Second, estimates of nuclear divergence and divergence times are tightly correlated ($r^2$=0.98; p-value = 0.01), an unsurprising result given the low estimates for migration during divergence ($M$ = 4$\times10^{-3}$ to 3.5$\times10^{-2}$ $\frac{migrants}{generation}$; Fig. 2B). Full model parameters are available in Table S4.

\subsection{Measuring reproductive isolation}
We used fine-scale spatial sampling and multilocus estimates of hybridization and introgression to infer the strength of RI. We sampled densely through each contact zone, averaging 20 individuals for each of 12 populations per contact ($N=55-406$; Table S1), with the geographic scale of sampling determined by preliminary data on the respective mtDNA transition (range of transect length $=2-16$ km). For \emph{S. basiliscus} N/\emph{S. lewisi}, we genotyped each lizard at six diagnostic, nuclear SNPs and one diagnostic, mitochondrial SNP using a PCR-RFLP approach. For all other lineage-pairs, we genotyped each lizard at ten diagnostic, nuclear SNPs and one diagnostic, mitochondrial SNP. Based on these genotypic data, we then calculated six indirect indices of RI (average nuclear cline width, mitochondrial cline width, coefficient of variance of nuclear cline width, Hardy-Weinberg disequilibrium ($F_{IS}$), linkage disequilibrium ($R_{ij}$), and percent hybrids) for each lineage-pair. Importantly, we note that these indices are independent measures of isolation, though some would show correlated responses under certain conditions, such as under a tension zone model \cite{BartonGale}.

We first describe general patterns at each contact zone before summarizing across all the zones, stepping from the least to most divergent contact (Fig. 3). In the \emph{L. coggeri} N/C contact, we see widespread introgression that extends throughout the sampled transect and evidence for two general patterns of introgression in nuclear loci: clines whose center and width is similar to the mtDNA cline and clines which show broad introgression of the Central (C) alleles into the Northern (N) lineage (Fig. 3A). In the \emph{S. basiliscus} N/C contact, we were unable to infer clines at all but one of the nuclear loci; it appears that northern alleles have almost completely introgressed into the Central lineage (Fig. 3B). The asymmetric hybridization in both the \emph{L. coggeri} N/C and \emph{S. basiliscus} N/C contacts could stem from stochastic, demographic or selective processes; disentangling the causes of asymmetry is not possible here so we focus on consensus patterns. Both the \emph{C. rubrigularis} N/S and \emph{L. coggeri} C/S show similar clines across all loci, and both show limited introgression beyond the contact zone (Fig. 3C and 3D). Finally, there is no evidence for hybridization between \emph{S. basiliscus} N and its ecomorphologically similar sister species \emph{S. lewisi}, even when sampled in sympatry (Fig. 3E).

Examining the correlation of divergence time with the six indices of RI, we see significant and strong correlations for all indices but mitochondrial cline width (Fig. 4). As predicted with increasing divergence time, we see decreased cline width and variance in cline width, fewer hybrids, and increased between- and within-loci disequilibria. These results are robust to our estimates of splitting times; using pairwise nuclear divergence gives quantitatively similar results (Fig. S4). Note that not all indices of RI could be estimated for all contacts. We did not infer cline indices for the \emph{S. basiliscus} N/\emph{S. lewisi} contact zone because of insufficient sampling, and we did not estimate either disequilibrium or hybridization measures for \emph{S. basiliscus} N/C as most nuclear loci were nearly monoallelic throughout the sampled contact zone.

To make our data more broadly comparable to other published data sets, we fit \emph{linear} and \emph{quadratic} models to the increase of RI through time \cite{Darters}. Although these models have no formal theoretical basis, they reflect the speed and accumulation at which total RI accumulates. We fit the models to the three indices -- $F_{IS}$, $R_{ij}$, and percent hybrids -- which are expected to be zero at the start of divergence. Using relative weights from AIC scores, we determined that total RI, as measured by each of these three indices, best fits a model of quadratic growth with time (Fig. S5A-C). Following \cite{Goubriere}, we then used our data to contrast three models for the accumulation of DMIs and found that our data fit the \emph{slowdown} model better than the \emph{linear} or \emph{snowball} models, suggesting the rate at which DMIs accumulated slowed down with time (Fig. S5D-F).

\section{Discussion}
By looking across five contacts in a clade of closely-related and ecologically-similar skinks in the Australian Wet Tropics, we find strong support for the prediction of increasing RI with divergence time. To our knowledge, this is the first comparative study of the strength of lineage boundaries across eco-morphologically similar lineages. These data support the view that phylogeographic splits of increasing depth can represent stages along the speciation continuum -- including genetically-cohesive lineages with long-term potential for persistence.

Interestingly, most other data sets comparing RI with divergence time show significantly more noise than ours \cite{Darters,Sasa}, even thogh these data sets were collected in controlled laboratory settings. In comparison, the strength of our correlations is unexpected, especially given that stochastic processes often influence hybrid zone structure and dynamics significantly \cite{Weta,TuckerEvol}. We speculate that the close fit between RI and divergence time in our study stems from the lack of overt divergent selection on eco-morphology, as varying strengths and forms of selection would be expected to introduce rate heterogeneity \cite{GavriletsBook}. In fact, the relationships between different indices of RI and divergence time are so strong that we can use them as a general metric for predicting the progress of speciation in this group. As has been suggested by \cite{KirkpatrickRavigne}, populations achieve "species status" when linkage disequilibrium (measured here as $R_{ij}$) is 0.5. Using this relationship, we find that our lineage-pairs are predicted to show $R_{ij} = 0.5$ at 7 Myr or 8.1 Myr after divergence (and at 0.79 or 0.81\% nDNA divergence), under a linear versus quadratic model for accumulation of RI, respectively. Although this specific calibration is contingent on our divergence time calibration and unlikely to be generalizable to other taxa, it does provide a yardstick for the tempo of speciation in this group.

Given this rate of speciation -- and noting that \emph{C. rubrigularis} N/S and \emph{L. coggeri} C/S show significant but incomplete isolation with even less time -- we suggest that the accumulation of RI here is rapid relative to a purely drift-driven model of the evolution of intrinsic RI via DMIs. A simplistic model of drift-driven accumulation of DMIs in allopatry suggests that the waiting time to speciation is approximately the number of substitutions needed for RI divided by the substitution rate \cite{GavriletsBook}, which, given the skinks' estimated mutation rate, could be on the order of hundreds of millions of years. Thus, accumulating substantial RI with minimal phenotypic divergence suggests (1) rapid drift-driven divergence along "holey adaptive landscapes" \cite{GavriletsBook}, (2) parallel selection driving mutation-order speciation \cite{FlaxmanNosil}, and/or (3) natural or sexual selection acting on more cryptic phenotypes, like chemiosensory production and perception. As yet, we lack the fine-scale ecological data necessary to characterize the barriers to gene flow acting in this group and to determine which of these hypothetical drivers of divergence are relevant \cite{SchemskeRI}. However, these data do confirm cryptic speciation among phylogeographic lineages and suggest that this could be common, in contrast to the present focus on speciation driven by divergent selection \cite{Schluter09}. 

Looking beyond the velocity of RI accumulation to its acceleration, we find that the total strength of RI increases exponentially through time in this clade (Fig. S5A-C). The pattern of exponentially increasing RI emerges when individual barriers to gene flow combine multiplicatively rather than additively \cite{Goubriere}, and it is occasionally recovered in other studies \cite{Mendelson2004}. This result suggests that as barriers to gene flow start to evolve, the cumulative effect of these barriers can grow quickly. Thus, species formation can be thought of as an accelerating process; particularly as RI decreases gene flow, which typically further promotes divergence \cite{Endler}. Further, although our data suggest the rate at which DMIs accumulate might decrease through time (Fig. S5D-F), we refrain from over-interpreting these results because our indices of RI potentially include both pre-zygotic and post-zygotic factors and few data have addressed the model's assumptions of equal and multiplicative fitness effects \cite{Goubriere}.

\subsection{Implications for cryptic speciation}
Because we can quickly and cheaply query geographic variation within species using mtDNA, deep mtDNA divergence is often used as an initial hypothesis for species delimitation. However, deep mtDNA divergence is neither necessary nor sufficient to delimit species \cite{MoritzBarcoding}, and it is often discordant with other genetic or phenotypic measures of divergence \cite{ToewsReview}. In this work, we provide additional evidence that mtDNA often presents an idiosyncratic perspective on historical dynamics, finding that mitochondrial cline width is the sole index of RI that had a non-significant correlation with divergence time. Given our and others' findings of mtDNA's idiosyncracy \cite{BallardReview}, we concur that the observation of deeply divergent mtDNA phylogroups is a useful start of taxonomic and phylogeographic studies, but it certainly should not be the end.

Further, since researchers have begun cataloguing diversity within species, most species have been found to show geographically-restricted variation \cite{AvisePhylo}. Why do some of these intraspecific lineages continuously diverge and exhibit reproductive isolation from sister-lineages, whereas others collapse \cite{SeehausenCollapse}? As suggested by many, geography is a powerful determinant \cite{AvisePhylo}. But, as we see here, RI can take millions of years to accumulate, particularly in the absence of strong ecologically-driven divergence, suggesting that isolation must be sustained across millions of years to lead to genetically independent lineages. In the absence of historical stability, geography can be insufficient, leading to the extensive introgression and discordance often reported in other systems \cite{McGuireCrota,DartersIncomplete}. Thus, as a working hypothesis, we suggest that climatically and geomorphologically stable regions, such as the major refugia of the AWT, are more likely to accumulate such cryptic diversity than are more spatio-environmentally dynamic regions. Indeed, broad-range introgression and discordance are exceedingly rare in the AWT, except amongst lineages endemic to the relatively unstable southern rainforest isolates \cite{saproSS,Bell2012}. 

Finally, these data add to a growing body of literature that support a Darwinian perspective on species formation \cite{MalletDarwin} and extend this perspective to cryptic diversification. Whether from lineages that likely diverged with gene flow \cite{MalletBMCEvo} or lineages that diverged in allopatry \cite{EnsatinaBMC}, these data sets show that the accumulation of RI is often a gradual process and that species are not static entities. Indeed, divergence is a continuous, reversible process \cite{SeehausenCollapse}. For these lineage-pairs, we find that the evolution of RI followed a predictable timeline during their divergence in allopatry (Fig. 4; Fig. S3). Now that lineages have expanded following the Last Glacial Maximum and the original barrier to gene flow (\emph{i.e.}, geography) has disappeared, these lineage-pairs will move again along the continuum.  For \emph{L. coggeri} N/C and \emph{S. basiliscus} N/C, this initial divergence will likely be reversed, because the lineage-pairs are hybridizing freely in the apparent absence of RI. However, \emph{L. coggeri} C/S and \emph{C. rubrigularis} N/S appear to be more strongly isolated, because the scale and extent of hybridization and introgression between these lineage-pairs are very limited. Given the limited eco-morphological differentation of these lineages, we hypothesis this RI is intrinsic and not environmentally-dependent and, thus, likely to maintain lineage boundaries even in changing environments. As such, these lineage pairs will likely continue to diverge. That said, RI between these lineages is not complete, but these "leaky" species boundaries can serve as a source of novelty, whether through the evolution of reinforcement or through the selective introgression of adaptive alleles \cite{Hoskin2005,Anderson}. 

\section{Methods}
\subsection{Sampling}
We sampled five contact zones in the AWT from 2008 to 2011 (Table S1): \emph{Carlia rubrigularis} N/S, \emph{Lampropholis coggeri} N/C, \emph{L. coggeri} C/S, \emph{Saproscincus basiliscus} N/C, and \emph{S. basiliscus} N/\emph{S. lewisi} (Fig. 1, Fig. S1). For each contact, we first identified the location of the contact zone by genotyping individuals at the mitochondrial genome (Fig. 1B). Then, we collected samples from populations geographically-isolated from the contact zone, which we used to infer demographic history and to develop markers. For four of the five contacts, we sampled individuals non-destructively along a linear transect through the contact zone  (Fig. S1). For \emph{S. basiliscus} N/\emph{S. lewisi}, we sampled opportunistically because initial data suggested the lineages were not hybridizing (Fig. S1). Data from the \emph{L. coggeri} C/S hybrid zone were previously published in \cite{GilliesHZ}, and we expanded the \emph{C. rubrigularis} N/S data set collected by \cite{Phillips2004}, by genotyping new genes, increasing sample sizes, and adding new populations.

\subsection{Morphological Analyses}
Adult lizards outside of the hybrid zones were measured at four standard characters for lizards: snout-vent length, head width, head length, and hind limb length. To analyze these data, we split each species data set by sex because these species show evidence of sexual dimorphism. Where deemed relevant by MANCOVA, we removed the effect of elevation using unnormalized residuals of morphological characters against elevation. We then conducted a scaled principal components analysis. Using the first two major orthogonal axes, we tested for morphological differentiation across phylogeographic lineages using MANOVA, conducted follow-up ANOVAs on results that were significant, and followed significant ANOVAs with Tukey HSD tests \cite{CRAN}.

\subsection{Genetic Data Collection}
We collected two types of genetic data: transcriptomic data from populations isolated from the contact zone and genotypic data from populations located in the transition zone between lineages. We collected transcriptomic data for five individuals per lineage; from these data, we created pseudo-reference assemblies and called individual genotypes as described in \cite{SinghalGenomics}. We used variants identified from the transcriptomic data to design PCR-RFLP markers; these markers were specific to each contact zone. We selected variants that were located in untranslated regions (UTRs) of genes and that were diagnostic for the two lineages. Marker details, including primers, annealing temperatures, and corresponding restriction enzymes, can be found in Table S2. In total, individuals in the \emph{S. basiliscus} N/\emph{S. lewisi} contact zone were genotyped at 6 nuclear markers and mtDNA, and individuals in the other four zones at 10 nuclear markers and their mtDNA (Table S1). All genotype data are available in the DataDryad package: XXX.

\subsection{Analysis}
To analyze the data, we characterized the demographic history of the lineages and calculated several indirect measures of RI. 
\subsubsection{Fitting divergence histories}
{\sffamily dadi} uses a diffusion approximation to fit a likelihood model for demographic history to the two-dimensional site frequency spectrum (2D-SFS). We inferred the 2D-SFS using {\sffamily ANGSD}, which is able to infer a population's SFS without calling individual genotypes \cite{Nielsen2012}. Here, we only used UTR sequence because UTRs are more likely to evolve neutrally than coding sequence \cite{Williamson}, and we restricted our analysis to high-coverage regions ($\ge$20$\times$) where we had greater confidence in genotype calling \cite{Nielsen2012}. To construct the unfolded SFS, we polarized SNPs with sequence data from other lineages in the clade. Inferred demographic parameters were converted from coalescent units to real-time units by using estimates of the nuclear mutation rate, assuming a molecular clock, and accounting for differences in total sequenced length across contacts \cite{Gutenkunst2009}. Our estimate of the nuclear mutation rate (9$\times10^{-10}$ $\frac{substitutions}{bp \times generation}$) is derived from fossil-calibrated estimates of the mitochondrial mutation rate in this broader clade of lizards \cite{BrandleySysBio} and estimates of the nuclear-mitochondrial substitution rate scalar as inferred from {\sffamily IMa2} results and \cite{saproSS}. For the \emph{S. basiliscus} N/\emph{S. lewisi} contact zone, because we did not have genomic data for \emph{S. lewisi}, we fit an isolation-with-migration model to previously published data \cite{saproSS} using {\sffamily IMa2} \cite{IMa2}. All {\sffamily  dadi} and {\sffamily IMa2} input files and related scripts are available in the DataDryad package: XXX.

\subsubsection{Measuring reproductive isolation}
To infer the strength of RI, we calculated six indices -- average nuclear cline width, mitochondrial cline width, coefficient of variance of nuclear cline width, Hardy-Weinberg disequilibrium ($F_{IS}$), linkage disequilibrium ($R_{ij}$), and percent hybrids. We first collapsed adjoining sampling localities into geographic populations based on their Euclidean distance to their nearest neighbor. Then, we fit clines to our data using the program {\sffamily Analyse} \cite{Analyse}. Second, using {\sffamily Analyse}, we calculated multilocus measures of Hardy-Weinberg and linkage disequilibrium for all geographic localities across all contacts. Third, we used {\sffamily Structure} to estimate each individual's hybrid index \cite{Structure} and {\sffamily NewHybrids} to calculate the number of hybrids in the contact zone \cite{NewHybrids}.

We used our data to contrast different models for how RI accumulates through time, including \emph{linear} and \emph{quadratic} models for the accumulation of total RI and \emph{linear}, \emph{snowball}, and \emph{slowdown} models for the accumulation of DMIs through time (c.f. \cite{Goubriere}). We restricted our analyses to three indices of RI -- $F_{IS}$, $R_{ij}$, and percent hybrids -- as starting values for these three indices could be predicted. We did all model fitting using the least-squares approach implemented in {\sffamily R} \cite{CRAN}, and we chose the best-fitting models by calculating the relative weight of each model based on AIC score.

\section{Acknowledgements}
For advice and discussions, we gratefully acknowledge C. Ellison, T. Linderoth, R. Pereira, and members of the Moritz Lab, and for comments on previous versions of this manuscript, we thank R. Damasceno, C. DiVittorio, J. Patton, R. Von May, and D. Wake. Additionally, T. Linderoth provided scripts used to infer the 2D-SFS from {\sffamily ANGSD} output. For assistance in the field, we thank A. Blackwell, E. Hoffmann, C. Hoskin, B. Phillips, M. Tonione, and S. Williams. Support for this work was provided by Museum of Vertebrate Zoology Annie Alexander Fund, National Geographic Society, NSF-GRFP and NSF-DDIG, and the SSE Rosemary Grant Award. Supercomputing resources used in this study were provided by the grid resources at the Texas Advanced Computing Center and Pittsburgh Supercomputing Center.

 \clearpage

\section{Figures \& Tables}

\begin{figure}[H]
\centering	  
\includegraphics{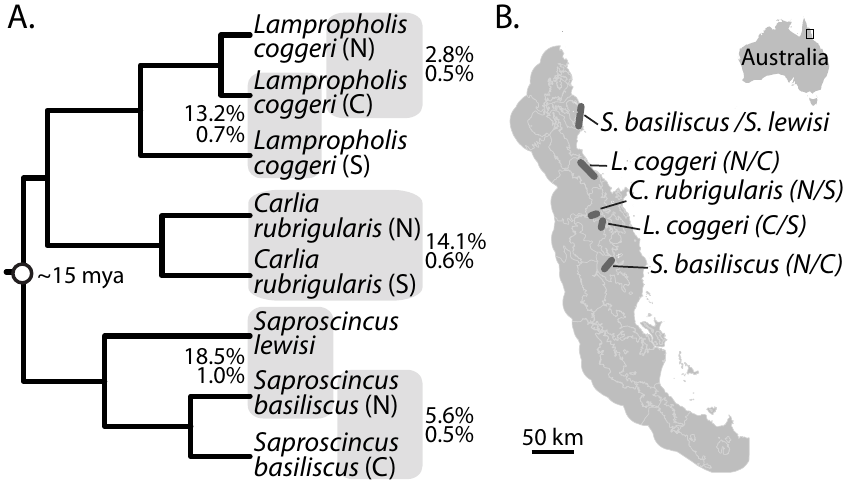}
	\caption{A. Phylogeny showing relationships among focal lineages (M. Brandley and C. Moritz, unpublished); boxes outline contact zones. Boxes are labeled with pairwise mitochondrial (top) and nuclear (bottom) divergence. B. A map of the Australian Wet Tropics, labeled with contact zones. A more detailed map of each contact zone is available in Fig. S1.}
\end{figure}

\begin{figure}[H]
\centering	  
\includegraphics{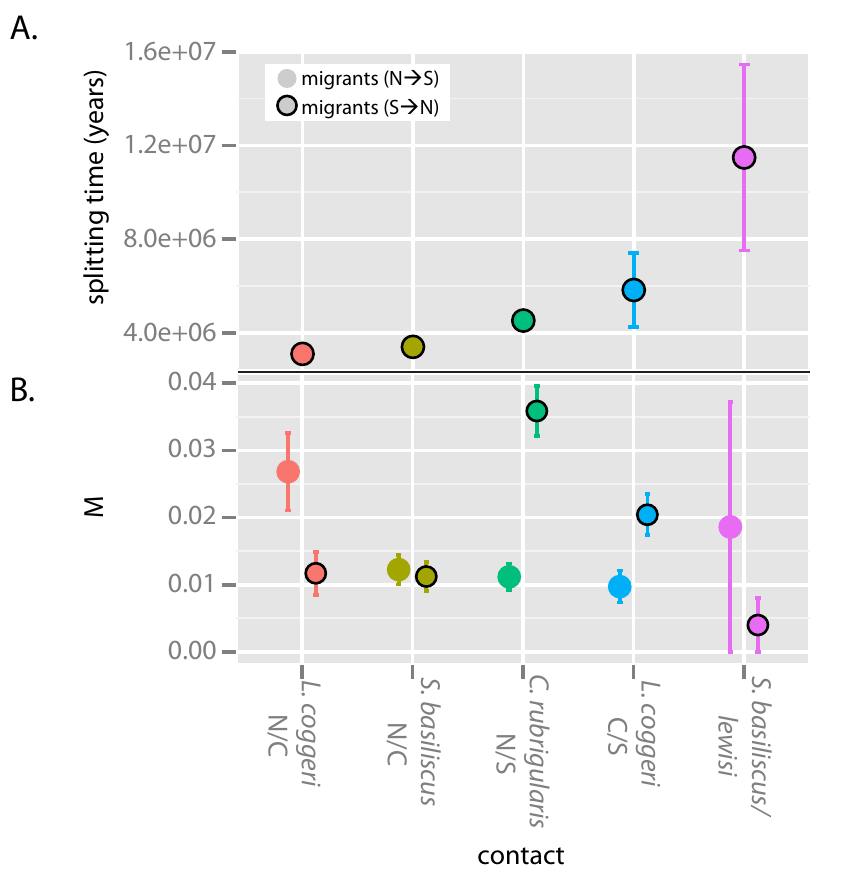}
	\caption{A. Divergence times and B. effective number of migrants for the lineage-pairs in this study, as inferred from the two-dimensional site frequency spectrum by {\sffamily dadi} and from {\sffamily IMa2}. Error bars reflect (as relevant) standard deviation or 95\% limits of posterior distribution. For \emph{Saproscincus basiliscus} N/\emph{S. lewisi}, \emph{S. lewisi} is the northern lineage and \emph{S. basiliscus} N is southern.}
\end{figure}

\begin{figure}[H]
\centering	  
\includegraphics{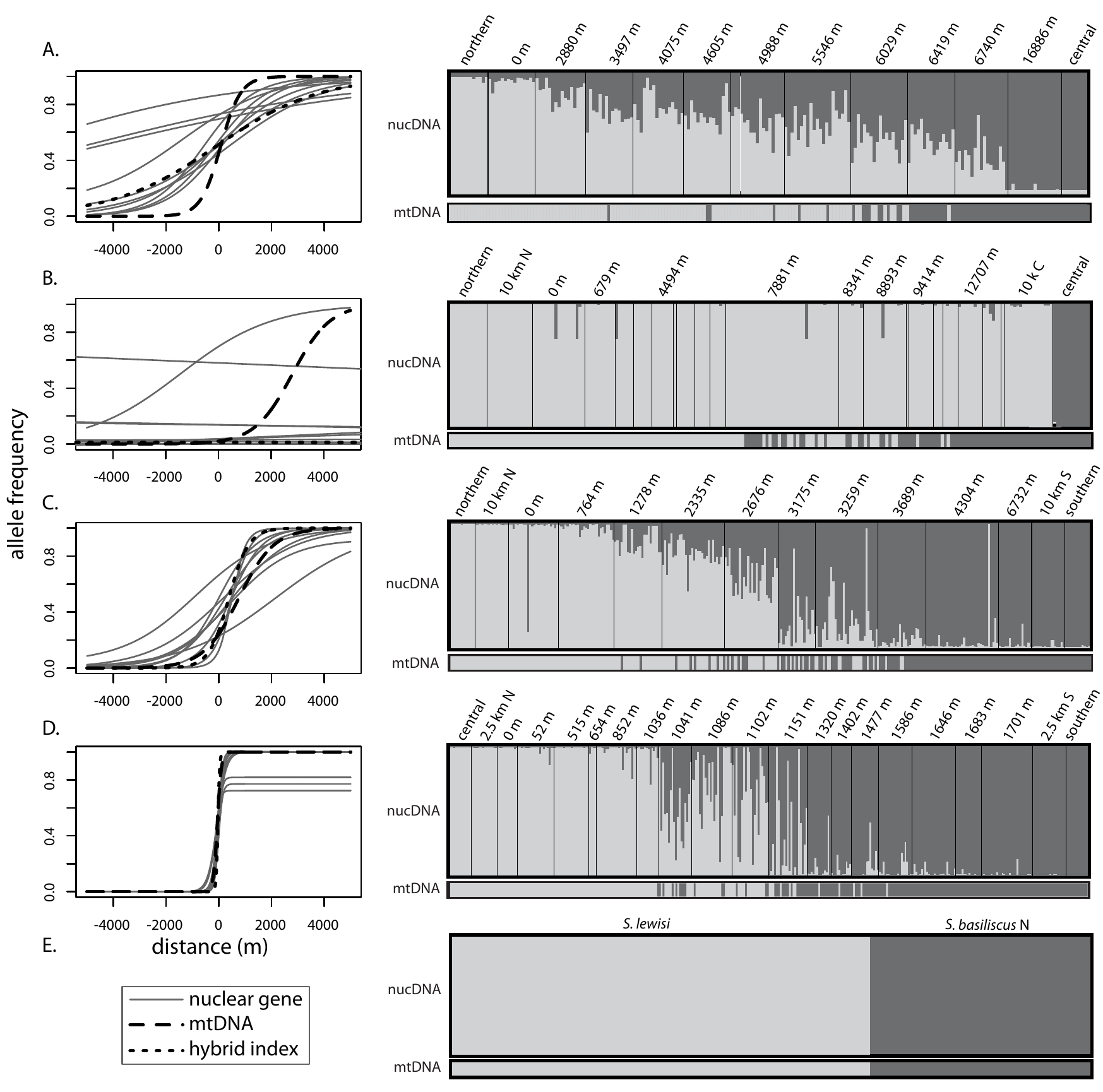}
	\caption{Cline fitting (left) and genetic clustering results (right) for contacts in the Australian Wet Tropics suture zone: A. \emph{Lampropholis coggeri} N/C, B. \emph{Saproscincus basiliscus} N/C, C. \emph{Carlia rubrigularis} N/S, D. \emph{L. coggeri} C/S, and E.\emph{S.basiliscus} N/\emph{S.lewisi}. For showing cline fitting results, distances along transects were recalculated so that each hybrid zone center was centered at 0 m. Scale for genetic clustering results differs among contacts.}
\end{figure}

\begin{figure}[H]
\centering	  
\includegraphics{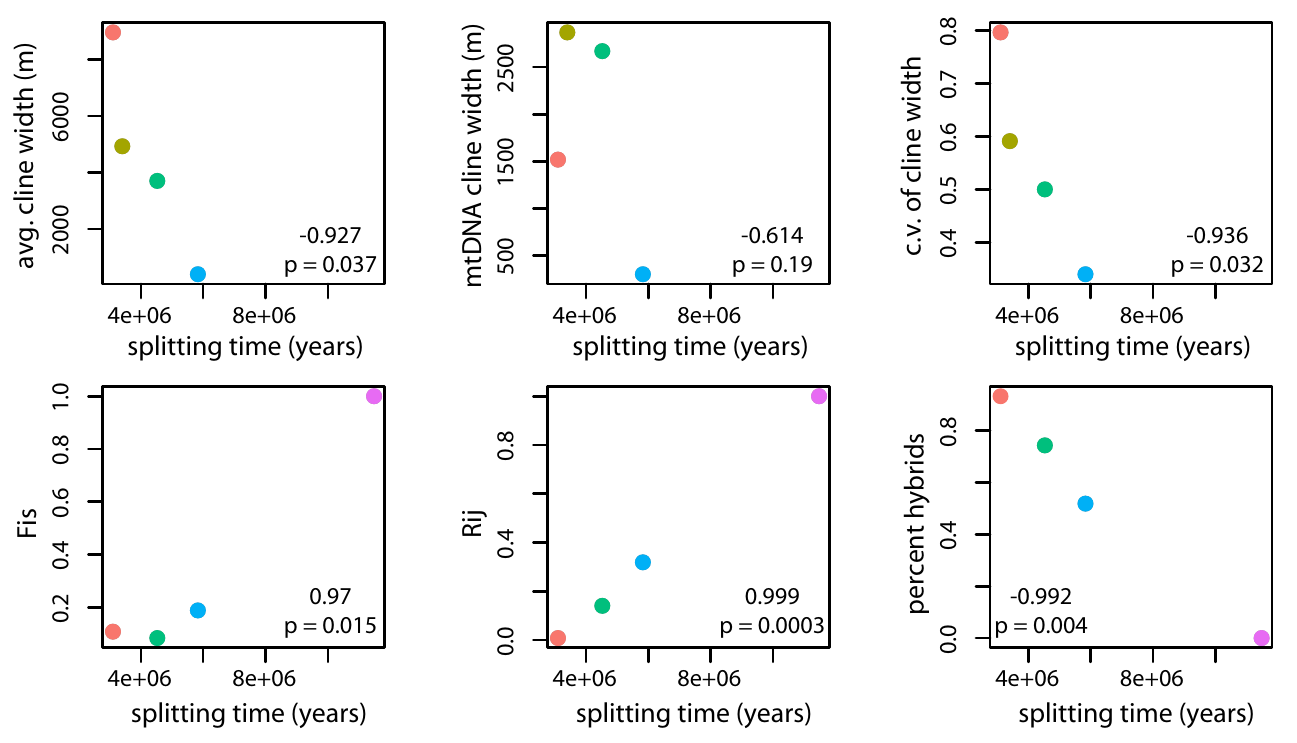}
	\caption{Comparative results showing the correlation between divergence time (as measured in years) and different indices of reproductive isolation: average nuclear cline width, mitochondrial cline width, coefficient of variance in nuclear cline width, Hardy-Weinberg disequilibrium ($F_{IS}$), linkage disequilibrium ($R_{ij}$), and percent of hybrids in the contact zone. Graphs are labeled with correlation coefficients. Colors follow Figure 2.}
\end{figure}

\setcounter{figure}{0}
\setcounter{table}{0}

\section{Supplemental Information: Figures \& Tables}

\begin{figure}[H]
\centering	  
\includegraphics[width=0.38\textwidth]{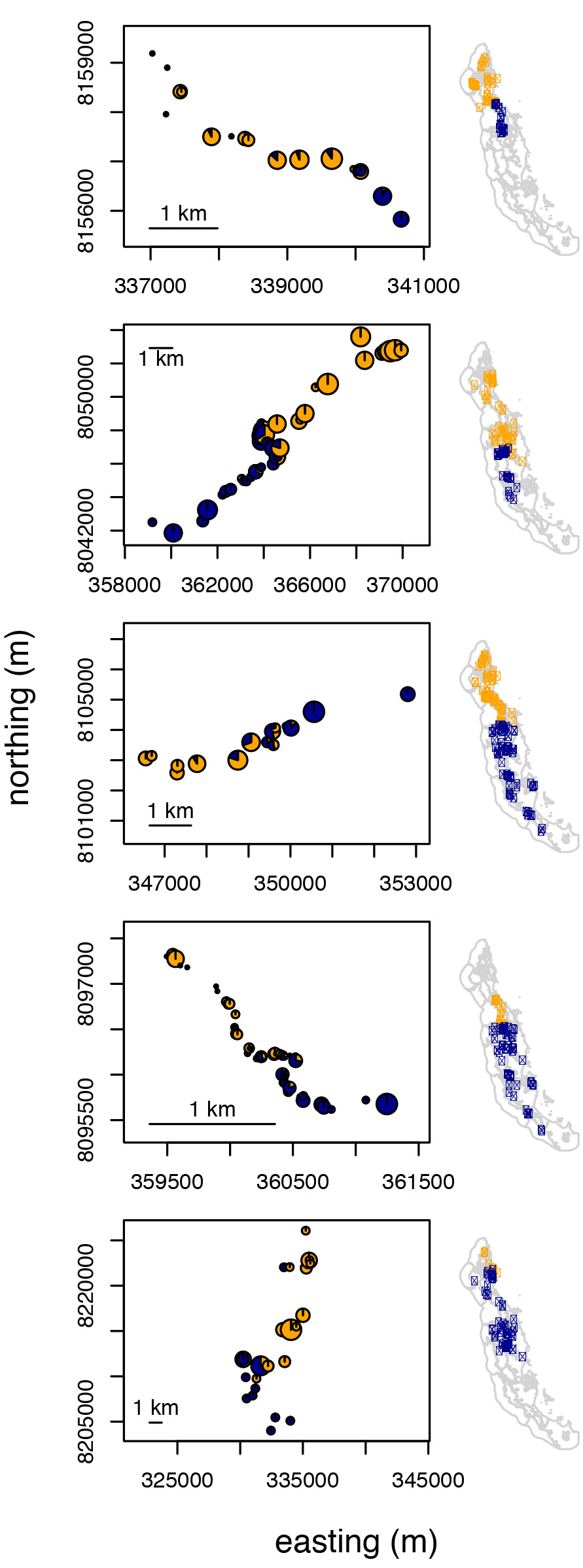}
	\caption{On left, transect for each contact zone, showing mitochondrial composition of unique localities with localities scaled according to sample size; on right, map of the Australian Wet Tropics showing the range of the phylogeographic lineages. From top to bottom, \emph{Lampropholis coggeri} N/C, \emph{Saproscincus basiliscus} N/C, \emph{Carlia rubrigularis} N/S, \emph{L. coggeri} C/S, and \emph{S. lewisi}/\emph{S. basiliscus} N.}
\end{figure}

\begin{figure}[H]
\centering	  
\includegraphics[width=0.8\textwidth]{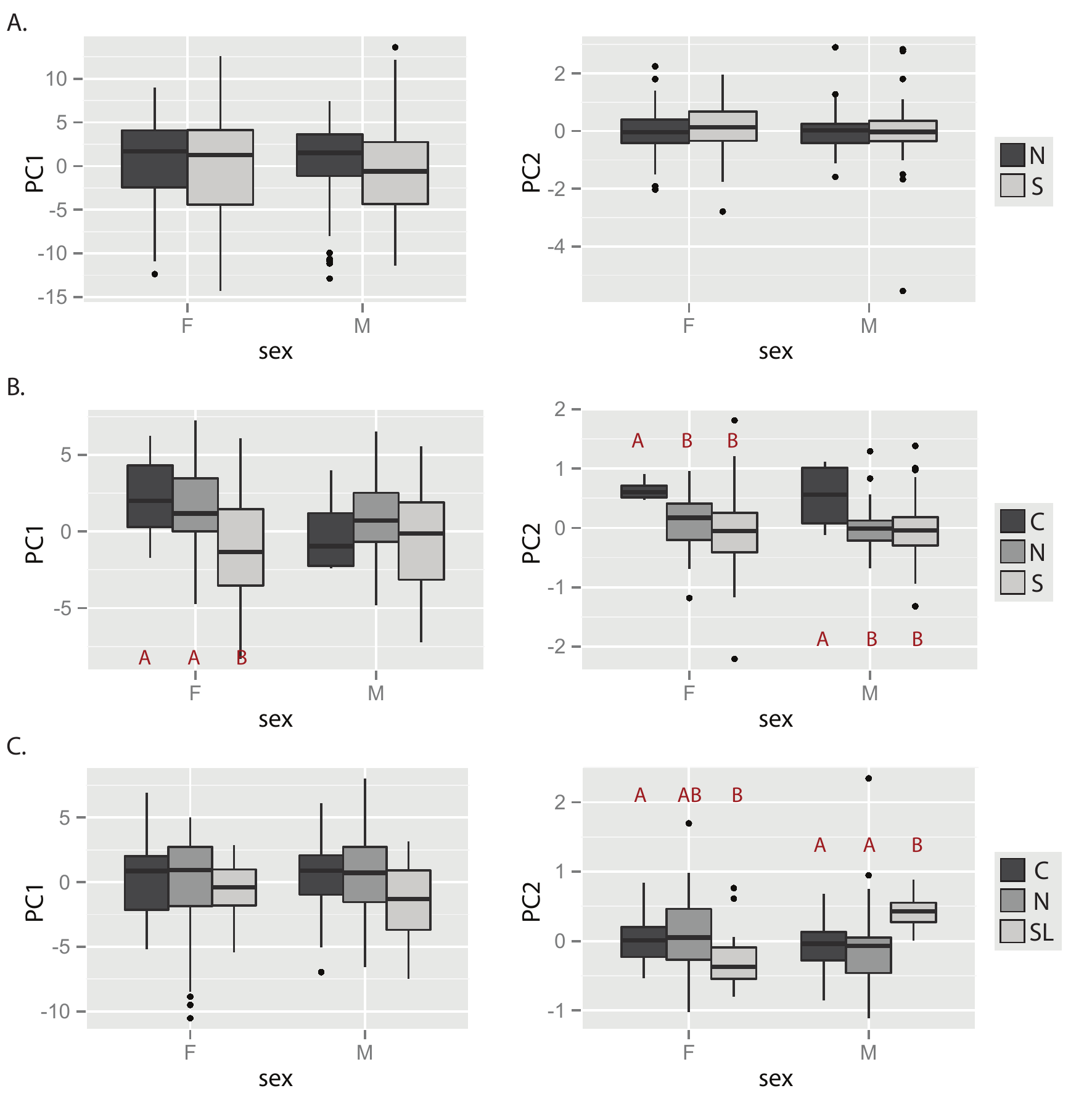}
	\caption{Morpological data summarized across sexes and across phylogeographic lineages within the four morphologically defined species in this study: A. \emph{Carlia rubrigularis} ($N_{\female}=223$, $N_{\male}=156$), B. \emph{Lampropholis coggeri} ($N_{\female}=174$, $N_{\male}=143$), and C. \emph{Saproscincus basiliscus} and \emph{S. lewisi} ($N_{\female}=119$,  $N_{\male}=119$). For each species, we present the first two axes of variation, as summarized by a principal components analysis. Significant differences are labeled in red.}
\end{figure}

\begin{figure}[H]
\centering	  
\includegraphics{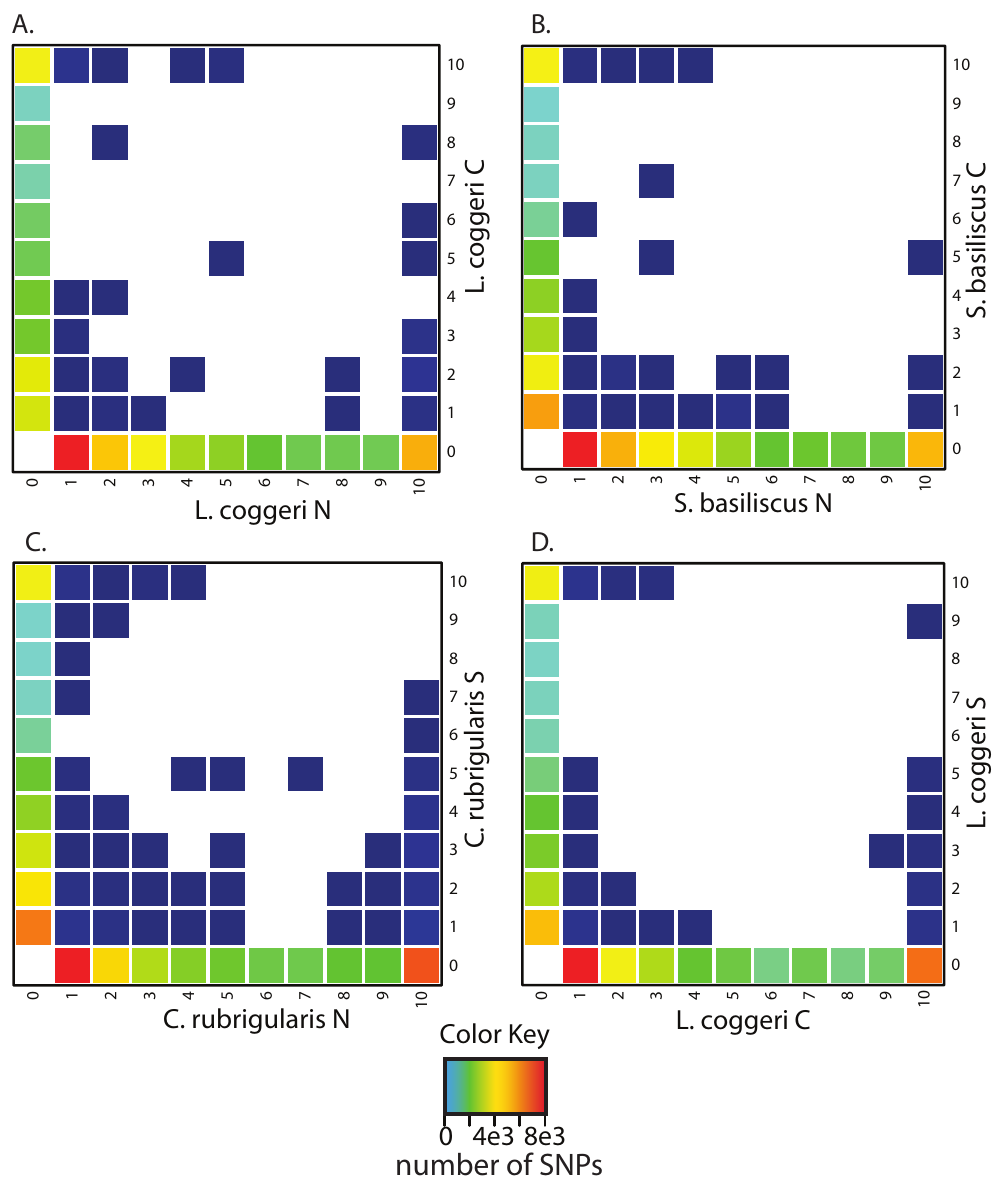}
	\caption{Two-dimensional site-frequency spectra (2D-SFS), as inferred by {\sffamily ANGSD}, for A. \emph{Lampropholis coggeri} N/C, B. \emph{Saproscincus basiliscus} N/C, C. \emph{Carlia rubrigularis} N/S, and D. \emph{L. coggeri} C/S. For each lineage-pair, we used a total of ten individuals, or twenty chromosomes, evenly split between the two lineages. Details on single nucleotide polymorphisms used to construct the 2D-SFS can be found in Table S2.}
\end{figure}

\begin{figure}[H]
\centering	  
\includegraphics[width=\textwidth]{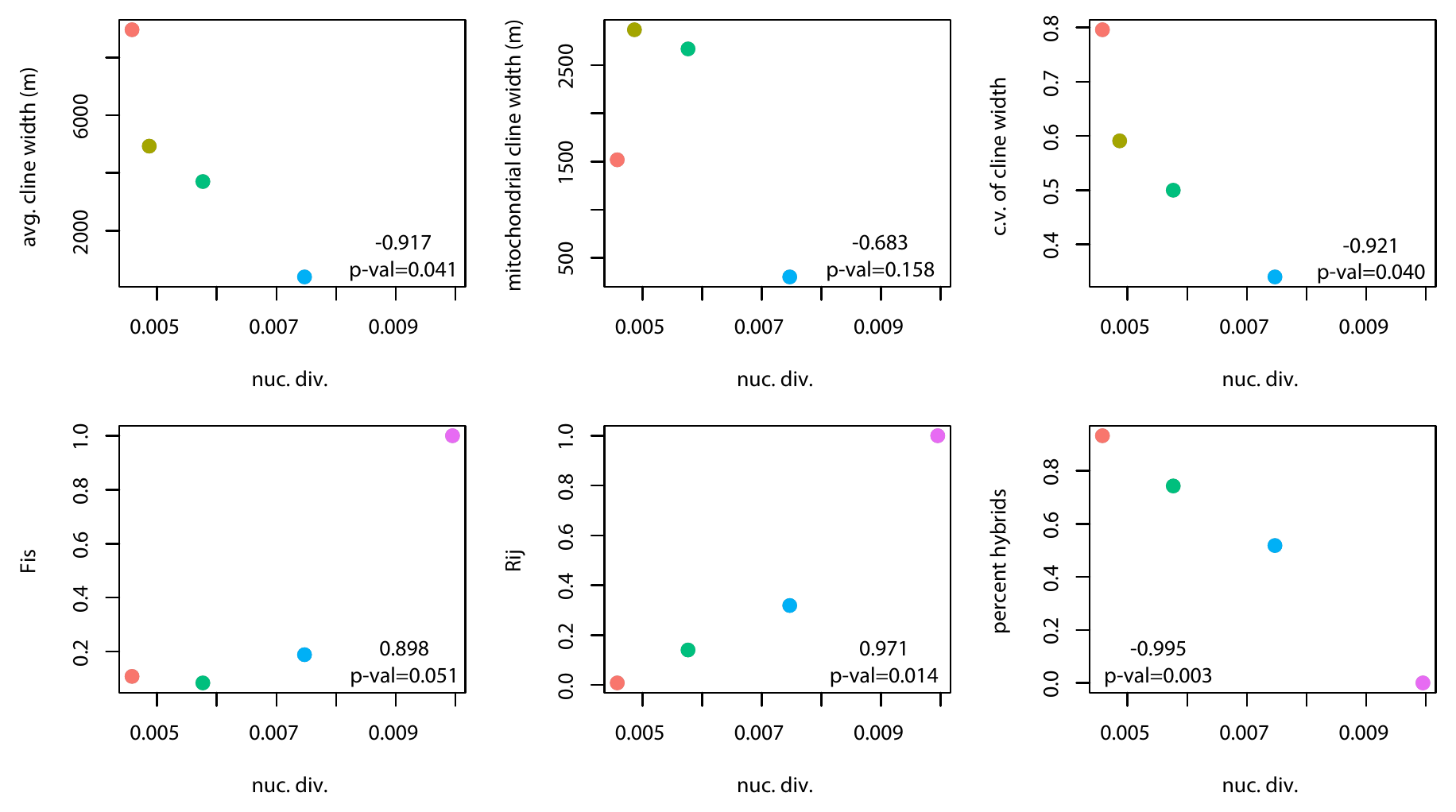}
	\caption{Comparative results showing the correlation between nuclear divergence and different indices of reproductive isolation: average nuclear cline width, mitochondrial cline width, coefficient of variance in nuclear cline width, Hardy-Weinberg disequilibrium ($F_{IS}$), linkage disequilibrium ($R_{ij}$), and percent of hybrids in the contact zone. Graphs are labeled with correlation coefficients.}
\end{figure}

\begin{figure}[H]
\centering	  
\includegraphics{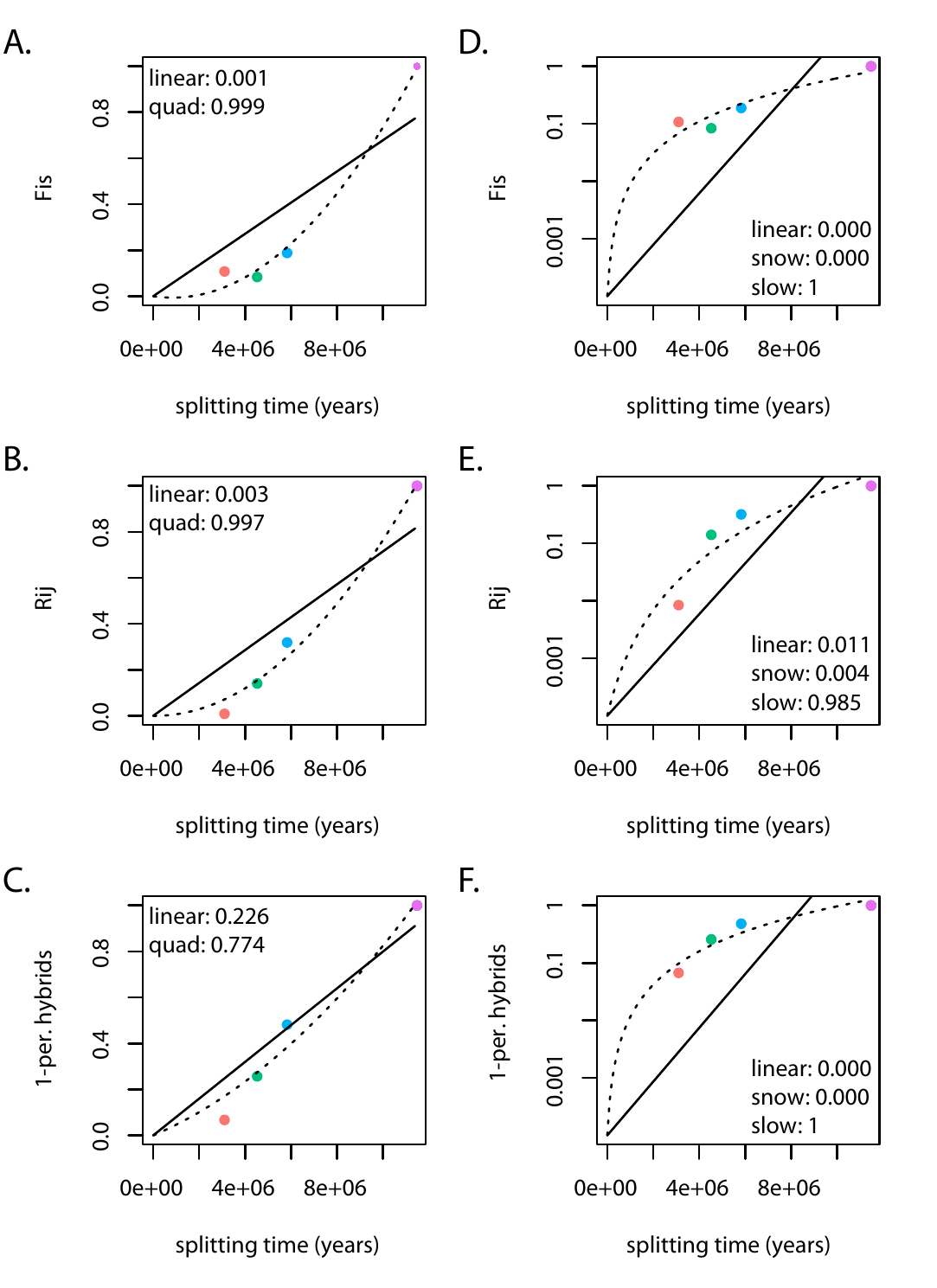}
	\caption{Model fitting for three indices of reproductive isolation. On the left, we fit linear (solid) and quadratic models (dotted) to the increase of reproductive isolation through time. On the right, we fit linear (solid), quadratic/snowball (solid), and slowdown models (dotted) to the log-linear increase of reproductive isolation through time [6]. Note that only one solid line is visible; model-fitting under the linear and snowball models gave the same result. Relative weights for the different models (as calculated via AIC scores) are shown for each model for each index. Colors follow Figure 2.}
\end{figure}

\begin{table}[H]
\centering
\label{graph}
\begin{tabular}{>{\centering\arraybackslash}m{4cm}  >{\centering\arraybackslash}m{2.5cm}   >{\centering\arraybackslash}m{3.5cm}  >{\centering\arraybackslash}m{2.5cm} }
\hline
contact zone & number of samples & number of transect populations & transect length \\ 
\hline
\emph{L. coggeri} N/C & 202  & 11 & 16 km \\
\emph{S. basiliscus} N/C & 209  & 10 & 12 km \\
\emph{C. rubrigularis} N/S & 308  & 10 & 7 km \\
\emph{L. coggeri} C/S & 406  & 17 & 2 km \\
\emph{S. basiliscus} N \& \emph{S. lewisi} & 55 & NA & 15 km \\
\hline
\end{tabular}
\caption{Sampling details for each contact zone. Transect populations are those used in estimation of clines.}
\end{table}

\begin{table}[H]
\centering
\includegraphics[angle=90,width=\textwidth]{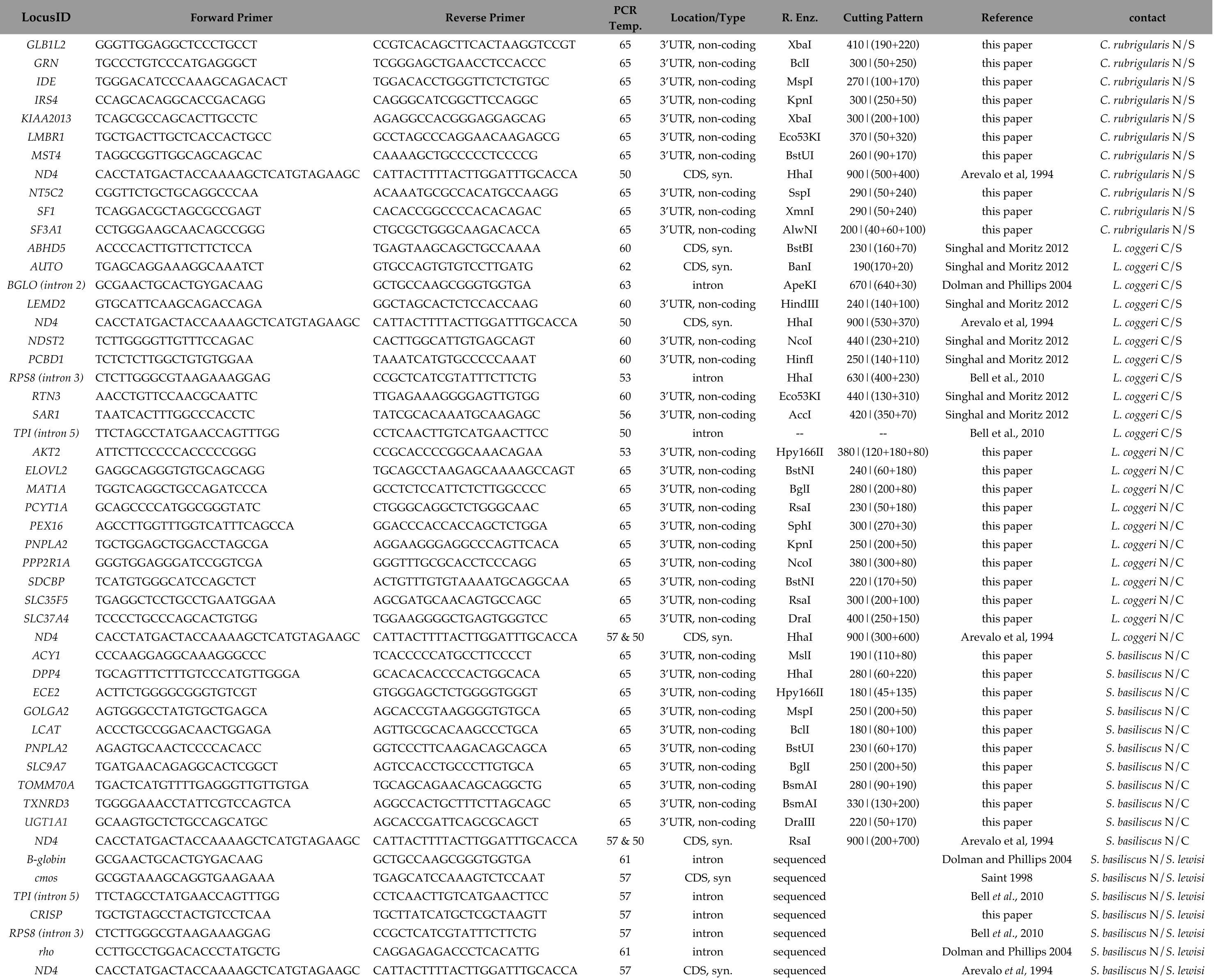}
\caption{The loci used in this study and their associated details.}
\end{table}

\begin{table}[H]
\centering
\label{graph}
\begin{tabular}{>{\centering\arraybackslash}m{4cm}  >{\centering\arraybackslash}m{2.5cm}  >{\centering\arraybackslash}m{2.5cm}  >{\centering\arraybackslash}m{2.5cm}  >{\centering\arraybackslash}m{2.5cm} }
\hline
contact zone & total number of SNPs & fixed SNPs & polymorphic SNPs & shared SNPs \\ 
\hline
\emph{L. coggeri} N/C		& 19884 & 3510 (17.7\%) & 16220 (81.6\%) & 154 (0.8\%) \\
\emph{S. basiliscus} N/C	& 29664 & 4712 (15.9\%) & 24798 (83.6\%) & 206 (0.7\%) \\
\emph{C. rubrigularis} N/S 	& 32264 & 6365 (19.7\%) & 25693 (79.6\%) & 369 (1.1\%) \\
\emph{L. coggeri} C/S 		& 41618 & 9260 (22.2\%) & 31989 (76.9\%) & 330 (0.8\%) \\
\hline
\end{tabular}
\caption{Details on the number of single nucleotide polymorphisms (SNPs), and their proportions, used in the two-dimensional site frequency spectrum (2D-SFS) for the contact zones analyzed with genomic data.}
\end{table}

\begin{table}[H]
\centering
\label{graph}
\begin{tabular}{>{\centering\arraybackslash}m{3cm}  >{\centering\arraybackslash}m{1cm}  >{\centering\arraybackslash}m{1cm}  >{\centering\arraybackslash}m{0.75cm}  >{\centering\arraybackslash}m{1.4cm} >{\centering\arraybackslash}m{1cm} >{\centering\arraybackslash}m{1cm} >{\centering\arraybackslash}m{1cm} >{\centering\arraybackslash}m{1cm} >{\centering\arraybackslash}m{1cm}}
\hline
contact zone & nuc. div. & mt. div. & theta ($\theta$) & div. time & $M_{12}$ & $M_{21}$ & $N_{1}$ & $N_{2}$ & $N_{A}$ \\
\hline
\emph{L. coggeri} N/C & 0.0046 & 0.028 & 3090 & 3.1 my & 0.0268 & 0.0117 & 408881 & 574006 & 1352453 \\
\emph{S. basiliscus} N/C & 0.0049 & 0.056 & 3644 & 3.4 my & 0.0123 & 0.0112 & 239822 & 919316 & 1352327 \\
\emph{C. rubrigularis} N/S & 0.0058 & 0.141 & 3775 & 4.5 my & 0.0112 & 0.0359 & 464585 & 1200178 & 1362782 \\
\emph{L. coggeri} C/S 	 & 0.0075 & 0.132 & 4608 & 5.8 my & 0.0097 & 0.0204 & 628695 & 1176557 & 2227376 \\
\emph{S. lewisi}/\emph{S. basiliscus} N & 0.0100 & 0.185 & NA & 11.4 my & 0.0186 & 0.0040 & 278501 & 740017 & NA \\
\hline
\end{tabular}
\caption{Parameter estimates for the isolation-with-migration model, as fit to the lineage-pairs. Populations labelled '1' are the northern lineage in each contact; populations labelled '2' the southern lineage.}
\end{table}

\end{document}